\documentclass[twocolumn,showpacs,10pt,aps,prd,amssymb,colordvi]{revtex4}

\newcommand{\Scri}{\mbox{$\cal J$}}

\usepackage{psfig}
\usepackage{epsfig}

\begin{document}

\title{Simulating the dynamics of relativistic stars via a light-cone
approach}

\author{Florian Siebel$^{1}$, Jos\'e A. Font$^{1,2}$, Ewald
M\"uller$^{1}$, Philippos Papadopoulos$^{3}$}
 
\address{$^{(1)}$Max-Planck-Institut f\"ur Astrophysik,
                 Karl-Schwarzschild-Str. 1,
                 D-85741 Garching, Germany \\
	 $^{(2)}$Departamento de Astronom\'{\i}a y Astrof\'{\i}sica,
		 Universidad de Valencia,
		 46100 Burjassot (Valencia), Spain \\
	 $^{(3)}$School of Computer Science and Mathematics,
                 University of Portsmouth, 
                 Portsmouth, PO1 2EG, UK}

\begin{abstract}
We present new numerical algorithms for the coupled Einstein-perfect
fluid system in axisymmetry. Our framework uses a foliation based on a
family of light cones, emanating from a regular center, and
terminating at future null infinity. This coordinate system is well
adapted to the study of the dynamical spacetimes associated with
isolated relativistic compact objects such as neutron stars.  In
particular, the approach allows the unambiguous extraction of
gravitational waves at future null infinity and avoids spurious outer
boundary reflections. The code can accurately maintain long-term
stability of polytropic equilibrium models of relativistic stars.  We
demonstrate global energy conservation in a strongly perturbed neutron
star spacetime, for which the total energy radiated away by
gravitational waves corresponds to a significant fraction of the Bondi
mass. As a first application we present results in the study of
pulsations of axisymmetric relativistic stars, extracting the
frequencies of the different fluid modes in fully relativistic
evolutions of the Einstein-perfect fluid system and making a first
comparison between the gravitational news function and the predicted
wave using the approximations of the quadrupole formula.
\end{abstract}

\date{\today}  

\pacs{04.25.Dm, 04.40.-b, 95.30.Lz, 04.40.Dg}

\maketitle

\section{Introduction}
\label{introduction}

The dynamics of relativistic compact objects is of interest both to
the astrophysical and relativity communities. These objects represent
on the one hand systems that may be active in stellar collapse,
central engines for Gamma Ray Bursts, and are potentially strong
sources of gravitational radiation.  On the other hand, they represent
non-trivial solutions to the Einstein equations, possibly exhibiting
directly the predictions of that theory for cosmic censorship and
black hole formation.

Despite the obvious interest in building a reliable spacetime
description of such systems, the process has been marred with
difficulties~\cite{StP85,Eva86}. One of the central problems concerns
the formulation and solution of the initial value problem for the
Einstein equations, a research effort that is still largely ongoing.
The numerical relativist must overcome a daunting list of problems,
including a long term stable formulation of the initial value problem,
the preservation of the constraints, the consistent imposition of
boundary conditions and the reliable extraction of physical
information (including gravitational waveforms).

We propose that for a wide and interesting class of spacetimes most of
the above concerns can be successfully dealt with by the use of a
characteristic initial value formulation for the Einstein
equations. In particular we argue that the study of the non-linear
dynamics of isolated relativistic stars is optimally performed within
this framework and we proceed to illustrate the technical procedures
that support this claim.

The incorporation of matter fields in the characteristic formulation
of the Einstein equation was considered as early as 1983~\cite{IWW83}
but the successful integration of the coupled system had to wait for
the development of stable algorithms for the {\em vacuum} Einstein
problem.  One dimensional schemes were developed
in~\cite{GWI92}. Algorithms for axisymmetric spacetimes, notably
including a regular origin, were presented in~\cite{GPW94}. Recently
using similar techniques, axisymmetric vacuum black hole spacetimes
have been evolved~\cite{Pap01}. Techniques for extending finite
difference algorithms to 3D were presented in~\cite{GLPW97}. Three
dimensional codes excluding the origin of the light cones (hence
presently unsuitable for stellar collapse studies) were presented in
\cite{BGLMW97}. For an alternative approach see~\cite{BaN00}.

With reliable algorithms for the vacuum Einstein equations available,
a new line of research for the incorporation of relativistic
hydrodynamics into numerical relativity was initiated
recently~\cite{PaF99}.  This approach brings into the considerations
the modern machinery from Computational Fluid Mechanics. In this
procedure, the evolution equations for the matter fields are solved
using relativistic high-resolution shock-capturing (HRSC)
schemes~\cite{Fon99,PaF99} based upon (exact or approximate) Riemann
solvers. A general formalism has been developed, which is now being
systematically applied to problems of increasing complexity.
Applications in spherical symmetry have already been presented in the
literature: investigations of accreting dynamic black holes can be
found in~\cite{PaF99,PaF002}. Studies of the gravitational collapse of
supermassive stars are discussed in~\cite{Lin00} and studies of the
interaction of scalar fields with relativistic stars are presented
in~\cite{SFP01}.  We note that there has been already a
proof-of-principle demonstration of the inclusion of matter fields in
three dimensions~\cite{BGLMW99}.

In this paper we apply the general framework laid out in~\cite{PaF99}
in the context of axisymmetric neutron star spacetimes. The code we
have developed has been built upon previous work by G\'omez,
Papadopoulos and Winicour~\cite{GPW94}, who constructed an
axisymmetric characteristic {\it vacuum} code.  The numerical
implementation of the field equations of general relativity is based
on the light cone formalism of Bondi~\cite{BBM62,Sac63} and
Tamburino-Winicour~\cite{TaW66}.

The broad target of this project is the investigation of relativistic
stellar dynamics (including collapse) using numerical relativity. A
prerequisite for such studies is the development of very accurate and
long-term stable general relativistic codes. It would seem that the
feature list of the characteristic approach makes it ideal for such
studies. One serious bottleneck of the approach is the breakdown of a
lightlike coordinate system in the emergence of light
caustics. Interestingly though, due to its quasi-spherical nature, no
matter how agitated a relativistic core, it is unlikely to focus the
light cones emanating from its interior.

The light cone approach has a number of advantages compared to
spacelike foliations: i)~It is physically motivated; the light cones
offer a simple and unambiguous physical gauge on which to base the
numerical spacetime grid. ii)~It is unconstrained; the evolved
variables capture rather directly the true degrees of freedom of the
gravitational field. iii)~It is very efficient; even in 3D, there are
but two partial differential equations to solve, along with a set of
radial integrations along the light cones.  iv)~It allows for well
defined compactification~\cite{Pen63} of the domain, which leads to
perfect outer boundary conditions. v)~Finally, and perhaps most
importantly, the above theoretical advantages have been shown in a
series of works to translate to remarkably numerically robust and
stable codes (see e.g.~\cite{GLMW98}). For a recent review of the
approach the reader is referred to~\cite{Win98}.

An important focus of research in numerical relativity has been the
investigation of dynamical spacetimes of relativistic stars. While the
pioneer investigations were mostly concerned with the study of
spherically-symmetric gravitational collapse scenarios the effort is
nowadays expanding to the study of the merging of compact neutron star
binaries. Advances in the numerical formulation of the equations as
well as in computational power have made possible to transform the
numerical evolution of neutron stars in general relativity into an
active field of research~\cite{SBS00,FSK99,SBS002,FDGS01,StF01}.

The paper is organized as follows: In Sec.~\ref{framework} we describe
our mathematical framework for characteristic numerical relativity in
axisymmetry. The next section describes our numerical implementation
in some detail. In Sec.~\ref{tests} we present tests to assess the
validity of the different regimes of our numerical implementation. In
Sec.~\ref{pulsations} we apply the current code in studies of stellar
pulsations. The last section concludes with a short summary.

\section{Mathematical framework}
\label{framework}

We will work with the coupled system of Einstein and
relativistic perfect fluid equations
\begin{eqnarray}
G_{ab} & = & \kappa T_{ab} \,, \\
\label{fluid}
\nabla_{a} T^{ab} & = & 0 \ ,\\
\label{continuity}
\nabla_{a} (\rho u^{a}) & = & 0,
\end{eqnarray}
with the latter two equations being the local conservation laws of
stress-energy and current density.  The energy-momentum tensor
$T_{ab}$ has the form
\begin{equation}
T_{ab} = \rho h u_{a} u_{b} + p g_{ab}.
\end{equation}
In this expression $\rho$ denotes the rest mass density, $h = 1 +
\epsilon + \frac{p}{\rho}$ is the specific enthalpy, $\epsilon$ is the
specific internal energy and $p$ is the pressure of the
fluid. The four-vector $u^{a}$, the 4-velocity of the fluid, fulfills 
the normalization condition $g_{ab} u^{a} u^{b} = -1$.  Using
geometrized units ($c=G=1$) the coupling constant is $\kappa = 8 \pi.$ In order
to close the system of fluid equations, an equation of state (EoS) has
also to be prescribed, $p=p(\rho, \epsilon)$.

\subsection{The Einstein equations for the Bondi metric} 

As a geometric and coordinate framework we will use the Bondi
(radiative) metric~\cite{BBM62} to describe the spacetime
\begin{eqnarray}
\label{Bondi}
& & ds^{2} = -\left(\frac{V}{r}e^{2 \beta} - U^{2} r^{2} e^{2 \gamma}\right) du^{2} 
 - 2 e^{2 \beta} du dr 
\nonumber \\ 
 &-& 2 U r^{2} e^{2 \gamma} du d\theta 
 + r^{2}
 (e^{2 \gamma} d\theta^{2} + e^{-2 \gamma} \sin^{2}\theta d\phi^{2}),\\
\nonumber
\end{eqnarray}
with coordinates $(x^{0},x^{1},x^{2},x^{3})=(u,r,\theta,\phi)$, where
$u$ is a null coordinate labeling outgoing light-cones, $r$ is the
radial coordinate whose level surfaces (two-spheres) have area $4 \pi
r^{2}$, and $\theta$ and $\phi$ are angular coordinates with $\phi$
being a Killing coordinate. The metric functions $V$, $U$, $\beta$ and
$\gamma$ depend on the coordinates $u$, $r$ and $\theta$. We choose the 
origin of the coordinate system $r=0$ to lie on the axis of our axisymmetric 
stellar configurations. With the above assumptions, the gravitational field
equations read explicitly
\begin{equation}
\label{EinsteinR}
R_{ab} = \kappa \left(\rho h (u_{a} u_{b} + \frac{1}{2} g_{ab}) - p g_{ab}\right),
\end{equation}
where the relevant components of the Ricci tensor $R_{ab}$ are
\begin{widetext}
\begin{eqnarray}
\label{Bondiequation}
\frac{r}{4}R_{rr} & = & \beta_{,r} - \frac{r}{2} (\gamma_{,r})^{2} \ ,
\\
\label{Uhyp}
2r^{2}R_{r\theta} & = & (r^{4} e^{2 (\gamma - \beta)} U_{,r})_{,r}  
       - 2 r^{2} (\beta_{,r\theta} - \gamma_{,r\theta} + 2 \gamma_{,r}
       \gamma_{,\theta} - \frac{2}{r} \beta_{,\theta} 
       - 2 \gamma_{,r} \ \cot\theta) \ ,
\\
\label{Vhyp}
-r^{2} e^{2 \beta} g^{AB} R_{AB} & = & 2 V_{,r} + \frac{1}{2} r^{4} e^{2(\gamma-\beta)} 
(U_{,r})^{2} - r^{2} U_{,r\theta} 
       - 4 r \ U_{,\theta} - r^{2} U_{,r} \cot\theta - 4 r \ U \ \cot\theta 
\\ \nonumber
       & & 
       + 2 e^{2(\beta-\gamma)} \{ -1-(3 \gamma_{,\theta} - \beta_{,\theta}) 
       \cot\theta - \gamma_{,\theta\theta} + \beta_{,\theta\theta} + 
       (\beta_{,\theta})^{2} + 2\gamma_{,\theta} (\gamma_{,\theta} - 
       \beta_{,\theta}) \} \ , 
\\
\label{gammaeq}
-r^{2} e^{2 \beta} g^{\phi\phi} R_{\phi\phi} & = & 2r (r \gamma)_{,ur} + 
(1-r\gamma_{,r}) V_{,r} - (r\gamma_{,rr} + \gamma_{,r}) V 
        - r(1-r\gamma_{,r}) U_{,\theta} - r^{2} (\cot\theta - 
	\gamma_{,\theta}) U_{,r} 
\nonumber \\
        & & 
        + e^{2(\beta-\gamma)}(-1-(3 \gamma_{,\theta} - 2 \beta_{,\theta}) 
	\cot\theta - \gamma_{,\theta\theta} 
        + 2\gamma_{,\theta} (\gamma_{,\theta} - \beta_{,\theta}))
\nonumber \\
        & &
	+ r \ U (2 r \gamma_{,r\theta} + 2 \gamma_{,\theta} + 
	r \gamma_{,r} \cot\theta - 3 \cot\theta) \ . 
\end{eqnarray}
\end{widetext}
In Eq.~(\ref{Vhyp})
$A,B$ denote the angular coordinates, $A,B=2,3$. As usual a comma
is used to denote a partial derivative.

The Einstein equations decompose into hypersurface equations,
evolution equations and conservation laws. The hypersurface equations,
Eqs.~(\ref{Bondiequation})-(\ref{Vhyp}), form a hierarchical set for
$\beta_{,r}$, $U_{,r}$ and $V_{,r}$. The evolution equation is an
expression for $(r\gamma)_{,ur}$ given by Eq.~(\ref{gammaeq}).  The
light-cone problem is formulated in the region of spacetime between a
timelike worldtube $\Gamma$, which in our case is located at the
origin of the radial coordinate $r=0$, and future null infinity
$\Scri^{+}$. Initial data $\gamma$ is prescribed on an initial light cone
$u=0$ in this domain. Boundary data for $\beta$, $U$, $V$ and $\gamma$ 
is also required on $\Gamma$.

As shown in the original paper of Bondi~\cite{BBM62}, the contracted
Bianchi identities for the vacuum field equations enforce all other
Ricci tensor components to vanish, if they vanish on a worldline.  In the
same way one can show, that the contracted Bianchi identities for the
matter system enforce all other components of the Einstein
equation~(\ref{EinsteinR}) (see~\cite{IWW83}).

The reformulation of the above form of the equations for numerical
integrations follows the work of Ref.~\cite{GPW94}. First, in order to
be able to compactify the entire spacetime and to better resolve
interesting parts of the numerical integration (e.g. a stellar
configuration centered at $r=0$) we allow, starting from a radial
coordinate $x \in[0,1]$, for a coordinate transformation of the radial
coordinate $x \to r(x)$, with an associated derivative 
\begin{equation}
dx/dr = f^2(x) \,.
\end{equation}  
This transformation generalizes the results of~\cite{GPW94}, where the
fixed grid $r(x) = x/(1-x)$ was used. Furthermore, in order to
eliminate singular terms at the poles we use the new
coordinate~\cite{Pap93}
\begin{equation}
y = -\cos \theta
\end{equation}
and we introduce the new variables
\begin{eqnarray}
S & = & \frac{V - r}{r^{2}} , \\
\label{Uhat}
\hat{U} & = & \frac{U}{\sin \theta}, \\
\nonumber\\
\label{hatgamma}
\hat{\gamma} & = & \frac{\gamma}{\sin^{2} \theta}.
\end{eqnarray}

The metric, hence, takes the form
\begin{eqnarray}
\label{Bondinumerics}
ds^{2} & = &\left(-\frac{V}{r} e^{2 \beta} + U^{2} r^{2} e^{2 \gamma}\right) du^{2} - 
2 f^{-2} e^{2 \beta} du dx  
\nonumber \\
 & & -2 \hat{U} r^{2} e^{2 \gamma} du dy  + r^{2} (e^{2 \gamma} \sin^{-2}\theta dy^{2} 
\nonumber \\
& & + e^{-2 \gamma} \sin^{2}\theta d\phi^{2}).
\end{eqnarray}

The hypersurface equations~(\ref{Bondiequation})-(\ref{Vhyp}) thus read
\begin{widetext}
\begin{eqnarray}
\label{bexp}
\beta_{,x} & = & \frac{r}{2} f^2 \bar{y}^{2}
(\hat{\gamma}_{,x})^{2} + \frac{r}{4} f^2 R_{xx} \ ,
\\
\label{Uexp}
\left(r^{4} f^2 e^{2(\hat{\gamma}\bar{y}-\beta)} \hat{U}_{,x}\right)_{,x} 
& = & 2 r^{2} \left\{ \beta_{,xy} - \frac{2}{r f^2} \beta_{,y} + 4 y \hat{\gamma}_{,x} 
+ \bar{y} [ 2 \hat{\gamma}_{,x}(\bar{y} \hat{\gamma}_{,y} - 2 y \hat{\gamma})-
\hat{\gamma}_{,xy} ] \right\} +2 r^{2} R_{xy} \ ,
\\
\label{Sexp}
r^{2} f^2 S_{,x} + 2 r S & = & -1 - 4 r y \hat{U} - r^{2} f^2 y 
\hat{U}_{,x} + 2 r \bar{y} \hat{U}_{,y}
+ \bar{y} \left(\frac{r^{2}}{2} f^2  \hat{U}_{,xy} - \frac{r^{4}}{4} 
f^{4} e^{2(\hat{\gamma}\bar{y}-\beta)} (\hat{U}_{,x})^{2}\right)
\nonumber \\
&   & - e^{2(\beta-\hat{\gamma}\bar{y})} \{ -1 - 12 \hat{\gamma} - 2 y \beta_{,y}
 + \bar{y} [10 \hat{\gamma} + 8 y \hat{\gamma}_{,y} + 8 \hat{\gamma}^{2} + 4 y 
 \hat{\gamma} \beta_{,y} + \beta_{,yy} + (\beta_{,y})^{2}] \\
\nonumber
&   & - \bar{y}^{2} [ 8 \hat{\gamma}^{2} + 2 \hat{\gamma}_{,y} \beta_{,y} + 
      \hat{\gamma}_{,yy} + 8 y \hat{\gamma} \hat{\gamma}_{,y}] 
      + 2 \bar{y}^{3} \hat{\gamma}_{,y}^{2} \} 
      - \frac{r^{2}}{2} e^{2 \beta} g^{AB} R_{AB} \ . 
\end{eqnarray}
\end{widetext}
We have used here the notation $\bar{y}=1-y^{2}$ and $()_{,x}$ and $()_{,y}$ to denote 
the partial derivatives with respect to $x$ and $y$, in contrast to the partial
derivatives with respect to $r$ and $\theta$. Note, that the Ricci tensor components 
are those in $x, y$-coordinates.

Due to this choice of variables, the Einstein equations are non-singular on the 
polar axis, where $y= \pm 1$. Note, that the $y$-component of the four-velocity fulfills
\begin{equation}
u_{y} = \frac{u_{\theta}}{\sin \theta},
\end{equation}
which is in analogy to Eq.~(\ref{Uhat}).

The evolution equation for $\hat{\gamma}$ is written in the form of a wave equation for the 
quantity $\hat{\psi}$
\begin{equation}
\hat{\psi} = r \hat{\gamma},
\end{equation}
\begin{equation}
\label{waveg}
\tilde{g}^{ab} \tilde{\nabla}_{a} \tilde{\nabla}_{b} \hat{\psi} = - e^{-2 \beta} \hat{H},
\end{equation}
where the $\tilde{~}$ quantities are induced by the 2-metric
\begin{equation}
d \sigma^{2} = -\frac{V}{r} e^{2 \beta}  du^{2} - 2 e^{2 \beta} du \ dr,
\end{equation}
i.e. explicitly
\begin{equation}
\tilde{g}^{ab} \tilde{\nabla}_{a} \tilde{\nabla}_{b} \hat{\psi} = - e^{-2 \beta} ( 2 \hat{\psi}_{ur} - (\frac{V}{r} \hat{\psi}_{,r})_{,r}),
\end{equation}
and
\begin{widetext}
\begin{eqnarray}
\nonumber
\hat{H} & = & - \frac{r}{2} f^2 \hat{U}_{,xy} - \hat{U}_{,y} - 
r f^2 \hat{\gamma}_{,x} \hat{U}_{,y} \bar{y} +  \frac{r^{3}}{4} 
f^{4} e^{2(\hat{\gamma} \bar{y}-\beta)} (\hat{U}_{,x})^{2} \\
\nonumber
&   & - \left(S + r f^2 S_{,x}\right) \hat{\gamma} + \frac{1}{r} 
e^{2(\beta - \hat{\gamma} \bar{y})} (\beta_{,yy} + (\beta_{,y})^{2}) + 4 \hat{\gamma} \hat{U} y \\
\nonumber
&   & + 2r f^2 \hat{\gamma} \hat{U}_{,x} y + 6r f^2 
\hat{\gamma}_{,x} \hat{U} y - \bar{y} \left(r f^2 \hat{\gamma}_{,y} 
\hat{U}_{,x} + 2r f^2 \hat{\gamma}_{,xy} \hat{U} + 2 \hat{\gamma}_{,y} 
\hat{U}\right) \\
\label{sourcewave}
&   & + \frac{1}{2r} e^{2 (\beta - \hat{\gamma} \bar{y})} \kappa
\rho h u_{y} u_{y} \ .
\end{eqnarray}
\end{widetext}

\subsection{The relativistic perfect fluid equations}

Whereas Eq.~(\ref{continuity}) is a strict conservation law,
Eq.~(\ref{fluid}) involves source terms when writing the
covariant derivatives in terms of partial derivatives.
In the presence of a Killing vector field it can be
recast as a conservation law. Following~\cite{PaF00}, the number 
of source terms in Eq.~(\ref{fluid}) is minimized using the 
equivalent form
\begin{equation}
\label{fluid2}
\nabla_{a} T^{a}_{~b} =  0.
\end{equation}

In our hydrodynamics code we use the form given by Eq.~(\ref{fluid})
in order to set up the evolution equation for the radial
momentum. This is motivated by stability considerations when evolving
spherical neutron star models (see below). However, to set up the
evolution equation for the polar component of the momentum we use
Eq.~(\ref{fluid2}).  This form of the conservation law eliminates
singular behavior of the y-component of the velocity at the polar axis.

After introducing the definitions $U^{0}=\sqrt{-g} T^{00}$,
$U^{x}=\sqrt{-g} T^{0x}$, $U_{y}=\sqrt{-g} T^{0}_{~y}$ and
$U^4=\sqrt{-g} \rho u^{0}$, the fluid equations can be cast into a first-order 
flux-conservative, hyperbolic system for the state-vector
${\bf U}=(U^{0},U^{x},U_{y},U^4)$
\begin{eqnarray}
\label{conservation}
\partial_{0} U^{0} + \partial_{j} F^{j0}    & = & S^{0} \ , \\
\partial_{0} U^{x} + \partial_{j} F^{jx}    & = & S^{x} \ , \\
\partial_{0} U_{y} + \partial_{j} F^{j}_{~y} \ & = & S_{y} \ , \\
\partial_{0} U^{4} + \partial_{j} F^{j4}    & = & S^{4}.
\end{eqnarray}
The flux vectors are defined as
 \begin{eqnarray}
 F^{j0} & = & \sqrt{-g} \ T^{j0} \ , \\
 F^{jx} & = & \sqrt{-g} \ T^{jx} \ ,\\
 F^{j}_{~y} & = & \sqrt{-g} \ T^{j}_{~y} \ ,\\
 F^{j4} & = &\sqrt{-g} \ \rho \ u^{j} \ ,
 \end{eqnarray}
 and the corresponding source terms read
\begin{eqnarray}
S^{a} & = & g^{ab} \ S_{b} + \sqrt{-g} T^{c}_{~b} \partial_{c}g^{ab} \
, \\
S_{a} & = &\sqrt{-g} \ T^{b}_{~c} \Gamma_{ab}^{~~c}
\nonumber \\ & = & -
\frac{\sqrt{-g}}{2} \rho h u_{c} u_{d} (g^{cd})_{,a} + p
(\sqrt{-g})_{,a} \ , \\
S^{4} & = & 0 \ ,
\end{eqnarray}

In the above expressions $\sqrt{-g}$ is square root of the four
dimensional metric determinant and $\Gamma^{a}_{bc}$ are the
Christoffel symbols.  

\subsection{Gravitational waves at null infinity}

Using a compactified coordinate $x \in [0,1]$, $\lim_{x \to 1} r(x) =
\infty$, we have future null infinity $\Scri^{+}$ on our grid, where
we can unambiguously extract waveforms.  Hence our approach does not
suffer from the problem most numerical relativity codes have to deal
with, i.e. extracting approximate gravitational waveforms at a finite
distance (for a comparison, see \cite{GoW92,IWW84}).  Performing a
power series expansion of $\gamma$ around future null infinity in
inverse powers of the radial coordinate $r$
\begin{equation}
\gamma = K + \frac{c}{r} + O(r^{-2}),
\end{equation}
the hypersurface equations~(\ref{Bondiequation})-(\ref{Vhyp}) yield an 
expansion~\cite{IWW83}
\begin{eqnarray}
\beta & = & H + O(r^{-2}),\\
U & = & L + O(r^{-1}),\\
V & = & r^{2}(L \sin\theta)_{,\theta}/\sin \theta + r
e^{2(H-K)}[1 + K_{,\theta \theta} \nonumber \\
\label{VM}
&   & + 2(H_{,\theta} \sin \theta)_{,\theta}/ \sin
\theta + 3 K_{,\theta} \cot\theta 
+ 4(H_{,\theta})^{2} 
\nonumber \\ 
& & - 4 H_{,\theta} K_{,\theta} - 2 (K_{,\theta})^{2}] - 2
e^{2H} M + O(r^{-1}),
\end{eqnarray}
where $M=M(\theta)$ denotes the Bondi mass aspect. As its
straightforward extraction at $\Scri^{+}$ will be spoilt by the
leading, diverging terms in Eq.~(\ref{VM}), we follow the procedure 
proposed in~\cite{GRWI93} to determine the Bondi mass of a numerical 
spacetime. We numerically solve the hypersurface equations for the new
metric variables
\begin{widetext}
\begin{eqnarray}
\label{tau}
2 \tau & = & (1-y^{2})^{-1/2}r^{3}e^{2(\gamma-\beta)}U_{,r} + 2
r \beta_{,y}
- r^{2}(1-y^{2})^{-1} e^{2 \gamma}[(1-y^{2}) e^{-2 \gamma}]_{,ry},\\
\label{mu}
2 \mu & = & - V + r^{2}[(1-y^{2})^{1/2}U]_{,y} + r^{3} e^{2 \beta}
\left[\frac{1}{2r}(1-y^{2})e^{-2 \gamma}\right]_{,yyr} + e^{2 \beta}
[(1-y^{2})e^{-2 \gamma} \tau]_{,y},
\end{eqnarray}
\end{widetext}
(see~\cite{GRWI93} for more details). 
The Bondi mass $M_{B}$ can then be readily computed as
\begin{equation}
\label{bondimass}
M_{B} = \frac{1}{4 \pi} \int \omega^{-1} e^{-2 H} \mu|_{x=1} \sin
\theta \ d \theta \ d \phi,
\end{equation}
where $\omega$ denotes the conformal factor relating the two-geometry
\begin{equation}
d \hat{s}^{2} = e^{2K} d \theta^{2} + \sin^{2} \theta e^{-2K} d \phi^{2}
\end{equation}
to the two-geometry of a unit sphere
\begin{equation}
d \hat{s}_{B}^{2} = d \theta_{B}^{2} + \sin^{2} \theta_{B} d\phi_{B}^{2}.
\end{equation}
The total energy emitted by gravitational
waves during the time interval $[u,u+du]$ in angular directions
$[\theta, \theta + d \theta] \times [\phi, \phi+d \phi]$ is~\cite{IWW83}
\begin{eqnarray}
dE &=& \frac{1}{16 \pi} \omega^{-1} e^{-2H} \left\{2 c_{,u} + \frac{(\sin\theta
\ c^{2} \ L),_\theta}{\sin\theta \ c} 
\right.
\nonumber \\ 
& & \left. + e^{-2K} \omega \sin \theta [\frac{(e^{2H}
\omega)_{, \theta}}{\omega^{2} \sin \theta}]_{, \theta} \right\}^{2} \sin
\theta d\theta d\phi du .
\end{eqnarray}
The quantities $K, c, H$ and $L$ are read off from the metric variables at
$\Scri^{+}$, e.g. $c = - (r^{2} f^2
\frac{d\gamma}{dx})|_{x=1}$ (our coordinate transformations $r=r(x)$
fulfill the requirement that $r^{2} f^2$ is finite).

For the extraction of waveforms seen from a distant inertial observer,
we have to transform our coordinate system to a Bondi coordinate
system. Following~\cite{BGLMW97} the Bondi coordinate time $u_B$ is
related to the retarded time $u$ as
\begin{equation}
du = \frac{1}{\omega} e^{-2H}du_B,
\end{equation}
whereas the angular Bondi coordinate $y_B=-\cos(\theta_B)$ can be calculated from
\begin{equation}
dy = \frac{1}{\omega^{2}} dy_B.
\end{equation}
With the definition of the news function
\begin{eqnarray}
\label{bondinews}
N &=& \frac{1}{2} \frac{e^{-2H}} {\omega^{2}}
\left\{2 c_{,u} + \frac{(\sin\theta
\ c^{2} \ L),_\theta}{\sin\theta \ c} 
\right .
\nonumber \\
& & \left . + e^{-2K} \omega \sin \theta [\frac{(e^{2H}
\omega)_{, \theta}}{\omega^{2} \sin \theta}]_{, \theta} \right\},
\end{eqnarray}
and after integrating over the angles $\phi=\phi_B$, one recovers the expression for
the total energy radiated according to Bondi~\cite{BBM62}
\begin{equation}
dE = \frac{1}{2} N^{2} dy_B \ du_B.
\end{equation}
This relation allows us to check global energy conservation 
\begin{equation}
\label{ec}
ec := M_{B}(u)-M_{B}(u=0) + \int_{0}^{u} \int_{-1}^{1} dE = 0.
\end{equation}

For the calculation of the Bondi mass, Eq.~(\ref{bondimass}), as well
as for the calculation of the Bondi news, Eq.~(\ref{bondinews}), we
have to determine the conformal factor $\omega$
\begin{equation}
d{\hat{s}}_{B}^{2} = \omega^{2} d{\hat{s}}^{2}.
\end{equation}
It can be shown that the coordinate $y_{B} =
- \cos(\theta_B)$ reads
\begin{eqnarray}
y_{B}(y) &=& \tanh \left( \frac{1}{2} \int_{-1}^{y}
\frac{e^{2K}-1}{1-\tilde{y}^2} d \tilde{y} + 
\frac{1}{2} \int_{1}^{y}
\frac{e^{2K}-1}{1-\tilde{y}^2} d \tilde{y}
\right.
\nonumber \\
& & \left. + \int_{0}^{y}
\frac{1}{1-\tilde{y}^2} d \tilde{y} \right).
\end{eqnarray}
The choice of the integration constants ensures regularity of $y_B$,
i.e. $lim_{y \to \pm 1} y_B = \pm 1$ and for spacetimes with
equatorial plane symmetry, $y_B$ is symmetric as well. 
The conformal factor can be written as
\begin{equation}
\label{conformalfactor}
\omega = \frac{2 e^{K}}{(1+y) e^{\Delta} + (1-y) e^{-\Delta}}, 
\end{equation}
where
\begin{equation}
\label{Delta}
\Delta(y) = \frac{1}{2} \int_{-1}^{y}
\frac{e^{2K}-1}{1-\tilde{y}^2} d \tilde{y} + \frac{1}{2} \int_{1}^{y}
\frac{e^{2K}-1}{1-\tilde{y}^2} d \tilde{y}.
\end{equation}
The regularity of the conformal factor can be directly seen if we write
Eq.~(\ref{Delta}) as
\begin{equation}
\label{Deltanum}
\Delta(y) = \int_{-1}^{y} d \tilde{y} \int_{0}^{1} d \alpha e^{2
\alpha K} \hat{K} + \int_{1}^{y} d \tilde{y} \int_{0}^{1} d \alpha e^{2
\alpha K} \hat{K},
\end{equation}
where using Eq.~(\ref{hatgamma}) we have defined
\begin{equation}
\hat{K} = \frac{K}{\sin^{2} \theta}.
\end{equation}

\section{The numerical implementation}
\label{implementation}

We use an equidistant grid covering our numerical domain $(x,y) \in
[0,1] \times [-1,1]$ with grid spacings $\Delta x = 1/N_{x}$, $\Delta
y = 1/N_{y}$, where $N_{x}+1$ is the number of grid points in the
radial direction and $2N_{y}+1$ is the number of grid points in
the angular direction ($N_y$ is the number of angular grid zones per 
hemisphere). All variables are defined on
the grid $(u^{n},x_{i},y_{j}) = (n \Delta u, i \Delta x, j \Delta y)$,
except for the quantities $\hat{U}$ and $\tau$ which are defined on a
staggered grid $(u^{n},x_{i+1/2},y_{j+1/2}) = (n \Delta u, (i+1/2)
\Delta x, (j+1/2) \Delta y)$. As shown by~\cite{GPW94} the use of a
staggered grid is necessary for stability.

\subsection{The fluid evolution}
\label{sec:fluid}

In our implementation of the fluid equations we closely follow the
work of Papadopoulos and Font~\cite{PaF99}.  The hyperbolic
mathematical character of the hydrodynamic equations allows for a
solution procedure based on the computation of (local) Riemann
problems at each cell-interface of the numerical grid. At cell $i,j$
the state-vector ${\bf U}$ is updated in time (from $u^n$ to
$u^{n+1}$) using a conservative algorithm
\begin{eqnarray}
{\mathbf U}_{i,j}^{n+1}={\mathbf U}_{i,j}^{n}
&-&\frac{\Delta u}{\Delta x}
(\widehat{{\mathbf F}}_{i+1/2,j}-\widehat{{\mathbf F}}_{i-1/2,j}) 
\nonumber \\
&-&\frac{\Delta u}{\Delta y}
(\widehat{{\mathbf G}}_{i,j+1/2}-\widehat{{\mathbf G}}_{i,j-1/2}) 
\nonumber \\
&+& \Delta u {\mathbf S}_{i,j} \,,
\end{eqnarray}
\noindent
where the numerical fluxes, $\widehat{{\mathbf F}}$ and
$\widehat{{\mathbf G}}$, are evaluated at the cell interfaces
according to some particular {\it flux-formula} which makes explicit
use of the full spectral decomposition of the system. For our
particular formulation of the hydrodynamic equations such
characteristic information was presented in~\cite{PaF99}.  We note
that unlike in previous work~\cite{SFP01}, we have now included the
metric determinant in the definition of the conserved quantities
$\bf{U}$. Nevertheless, as was explicitly demonstrated
in~\cite{PaF00}, the eigenvalues needed for the computation of the
local Riemann problems remain unchanged.

In more precise terms the hydrodynamics solver of our code uses a
second order Godunov-type algorithm, based on piecewise linear
reconstruction procedures at each cell-interface~\cite{vanleer2,vanleer} and
for simplicity the HLLE approximate Riemann solver~\cite{harten,einfeldt}.  
For the time update we use the second order Runge-Kutta algorithm
derived in~\cite{ShO89}. General information on HRSC schemes in
relativistic hydrodynamics can be found, e.g. in~\cite{Fon99,Mue98}
and references therein.

We use the procedure described in~\cite{PaF00} to explicitly recover
the physical (primitive) variables for the perfect fluid EoS
$p = (\Gamma-1) \rho \epsilon$, $\Gamma$ being the adiabatic index of
the fluid. We solve for the primitive variables $\epsilon$, $\rho$,
$u^x$ and $u_y$ and then use the normalization condition to determine 
$u^{u}$.

\subsection{The metric evolution}

\begin{figure}[t]
  \begin{center}
    \begin{picture}(0,0)%
	\includegraphics{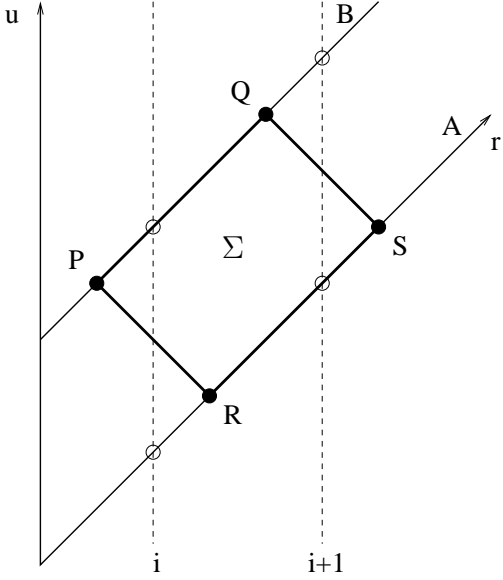}%
    \end{picture}%
    \setlength{\unitlength}{2329sp}%
	\begingroup\makeatletter\ifx\SetFigFont\undefined%
	\gdef\SetFigFont#1#2#3#4#5{%
	  \reset@font\fontsize{#1}{#2pt}%
	  \fontfamily{#3}\fontseries{#4}\fontshape{#5}%
	  \selectfont}%
	\fi\endgroup%
	\begin{picture}(5282,6097)(2026,-7636)
	\put(4351,-4261){\makebox(0,0)[lb]{\smash{\SetFigFont{12}{14.4}{\familydefault}{\mddefault}{\updefault}{$\Sigma$}%
}}}
     \end{picture}
    \caption{Parallelogram $PQRS$ consisting of ingoing and outgoing light
    cones. We have displayed the outgoing light cones at two different
    coordinate values $u$, $A$ and $B$,  foliating our spacetime. 
    For simplicity, all null lines have angles
    $45^{\circ}$. In addition to the parallelogram we have marked
    radial grid points at $i$ and $i+1$.
    \label{parallelogram}}
  \end{center}
\end{figure}

For the time update of the metric field $\hat{\gamma}$, we solve the
wave equation~(\ref{waveg}) with a so-called {\it parallelogram
algorithm}, based on a parallelogram consisting of ingoing and
outgoing characteristics~\cite{GWI92,GPW94,Win00}. We closely follow
the implementation of~\cite{GPW94} to which the reader is referred for
more details. In contrast to this work, however, we only use a single form of
the wave equation in our numerical implementation, which is regular at
$\Scri^{+}$.  This regularization is accomplished by rewriting the
parallelogram identity in terms of the quantity
\begin{equation}
\Phi = \hat{\psi} f.
\end{equation}
Explicitly, for a parallelogram $\Sigma$ with left upper corner $P$, right
upper corner $Q$, left lower corner $R$ and right lower corner $S$
(see Fig.~\ref{parallelogram}) the algorithm reduces to 
\begin{eqnarray}
\nonumber
\Phi_{Q} & = & \frac{1}{4} \ \Delta u \ r_{Q}
	 f|_{Q} \ \hat{H}_{c}\\
\nonumber
	 & + &
	 \frac{f|_{Q}}{f|_{P}}(\Phi_{P}-\frac{1}{4}
	 \ \Delta u \ r_{P} f|_{P} \ \hat{H}_{c})\\ 
\nonumber
	 & + &
	 \frac{f|_{Q}}{f|_{S}}(\Phi_{S}+\frac{1}{4}
	 \ \Delta u \ r_{S} f|_{S} \ \hat{H}_{c})\\
	 & - &
	 \frac{f|_{Q}}{f|_{R}}(\Phi_{R}+\frac{1}{4}
	 \ \Delta u \ r_{R} f|_{R} \ \hat{H}_{c}).
\end{eqnarray}
In the above expression
$\hat{H}_{c}$ denotes the source term of Eq.~(\ref{sourcewave}) at the center
of the parallelogram approximated to second order accuracy. The
regularity of this expression can be easily seen for our compactified
grids
\begin{equation}
r(x) = a \frac{x}{1-x^{b}}, \hspace{0.5cm} b \ge 1,
\end{equation}
as 
\begin{equation}
r f = \sqrt{a} \frac{x}{\sqrt{1+(b-1)x^{b}}}
\end{equation}
and 
\begin{equation}
\label{regs}
\frac{f|_{Q}}{f|_{S}} =
\frac{1-x_{Q}}{1-x_{S}} \ \frac{\sum_{i=0}^{b-1}
x_{Q}^{i}}{\sum_{j=0}^{b-1} x_{S}^{j}} \  \frac{\sqrt{1+(b-1)x_{P}^{b}}}{\sqrt{1+(b-1)x_{Q}^{b}}}.
\end{equation}
All the terms on the right hand side of Eq.~(\ref{regs}) 
are explicitly regular for $x \to 1$ ($r \to \infty$), 
except for the first factor which is regular according to 
the discussion in~\cite{GPW94}. 

We discretize the hypersurface equation~(\ref{bexp}) (and
similarly Eq.~(\ref{Sexp})) as
\begin{equation}
\beta_{i,j} = \beta_{i-1,j} + {\cal{H}_{\beta}}_{i-1/2,j} \Delta x,
\end{equation}
where $\cal{H}_{\beta}$ denotes the right hand side of Eq.~(\ref{bexp}).
To solve the hypersuface equation~(\ref{Uexp}), we discretize the
alternative equation
\begin{eqnarray}
2x f \left(r^{4} f^{4} 
\hat{U}_x\right)_{,x^{4}} + 
r^{2} f^2 \left( -\frac{1}{2} \left( f^2 \right)_{,x} 
\frac{r^{2}}{x^{2}} f^2\right)^{\frac{1}{2}}
\nonumber
\\
- \frac{r^{2}}{x^{2}} \left(f^2\right)^{\frac{3}{2}} 
\left(\beta_{,x} - {\hat{\gamma}}_{,x} \bar{y} \right) \hat{U}_{,x}
 = 
 \frac{r^2}{x^{2}}
f^3  e^{2(\beta-\bar{y}\hat{\gamma})} \cal{H}_{U},
\end{eqnarray}
where the right hand side of Eq.~(\ref{Uexp}) has been denoted by $2r^{2}
\cal{H}_{U}$. The derivative
$\partial/\partial x^{4} = \frac{1}{4x^{3}} \partial/\partial x$ was
introduced to ensure regularity at the origin. 

In the following, we describe the order of the time update from
light-cone $A$ at time $u$ to light-cone $B$ at time $u+\Delta u$ (see
Fig.~\ref{parallelogram}). Let us assume that we know the primitive and
conserved fluid variables, and the metric quantities $\hat{\gamma}$,
$\beta$, $\hat{U}$ and $S$ on the light cone $A$. In a first step we
globally determine the conserved fluid variables on $B$. For the
metric update, in contrast, we march from the origin to the exterior
of the light-cone $B$. Having previously obtained the variables up to
grid point $i$ (either from the specific boundary treatment at the origin 
or during the marching process), we first determine $\hat{\gamma}$ at 
grid point $i+1$. In a second step we solve for $\beta$, $\hat{U}$ and $S$ at
$i+1$ in that particular order, recovering the primitive variables with the
metric thus obtained. As the hypersurface integration for the metric
depends on the primitive variables at the grid point to be determined,
we iterate the hypersurface and recovery algorithm until convergence.

In order to obtain stability in our explicit algorithms when solving the 
fluid and metric equations we have to fulfill the Courant-Friedrichs-Levy
condition - the numerical domain of dependence must include the
analytical domain of dependence. This limits the maximal time step
allowed in each time update. Calculating the characteristic speeds for
the fluid system, the fluid update sets a limit on the time step as
\begin{equation}
\label{timehydro}
\Delta u \le \min(c_{1} \Delta x, c_{2} \Delta y),
\end{equation}
where $c_{1}$ and $c_{2}$ are constants and the minimum is calculated for
the entire fluid grid. 
For the metric update, it can be shown that the
evolution near the origin sets the stricter theoretical limit 
\begin{equation}
\label{timemetric}
\Delta u = c_{3}  \Delta r (\Delta y)^{2},
\end{equation}
with $c_3 = 0.5$.
In numerical experiments we found, however, that with our coupled
code $c_{3} \approx 10$, in good agreement with the result
$c_{3} = 8$ reported in~\cite{Pap93}. 
With this result the time step restriction from
Eq.~(\ref{timemetric}) is not much stronger than the time step restriction from
Eq.~(\ref{timehydro}), at least for the angular resolutions we can afford.
 In the simulations described in
this paper we use time steps of 0.6 times the maximal time step consistent
with the fluid evolution Eq.~(\ref{timehydro}). 
With this procedure, it was not necessary to
implement implicit methods for the metric update.

As for the vacuum equations the origin of
coordinates, where we assume that the coordinate system is a local
Fermi system, needs special care. The main change in the falloff
behavior of the metric variables due to the presence of
the fluid is in $\beta$. We impose a falloff behavior of the 
metric field
$\hat{\gamma}$ as
\begin{equation}
\label{origingamma}
\hat{\gamma} = a \ r^{2} + b \ r^{3},
\end{equation}
where $a$ and $b$ are constants. 
At the origin the radial dependence of $\beta$ is $\beta = O(r^{2})$, 
instead of $\beta = O(r^{4})$ for the vacuum case (from
Eq.~(\ref{bexp})).  
To ensure regularity  of the fluid at the origin, we have to impose~\cite{IWW83}
\begin{eqnarray}
u_{r} & = & D + Ey + O(r), \\
u_{y} & = & Er + O(r^{2}),
\end{eqnarray}
where $D$ and $E$ fulfill $-D^{2}+E^{2}=-1$. With the definitions (the
leading terms can be extracted from the hypersuface equations)
\begin{eqnarray}
\beta & = & \frac{r^{2}}{8} \kappa \rho h (D+Ey)^{2} + F(y) r^{3},\\
\label{originU}
\hat{U} & = & 4 y (a r + \frac{3}{5} b r^{2}) + \frac{1}{2} \kappa \rho h
(D+Ey) Er \nonumber \\ &&+ C(y) r^{2}, \\
\kappa \rho h u_{y} u_{y} & = & \kappa \rho h E^{2} r^{2} + G(y) r^{3},
\end{eqnarray}
the quadratic terms in $r$ in the wave equation reduce to an equation for $a$
\begin{equation}
\label{waveorigin}
a_{,u} = \frac{6}{5} b - \frac{1}{3}C_{,y} +
\frac{1}{6}F_{,yy}+\frac{1}{12} G.
\end{equation}
(In the absence of matter this reduces to $a_{,u} = \frac{6}{5} b$,
see~\cite{GPW94}). We extract $a$, $b$, $C$, $F$ and $G$ 
at the old time slice and then solve
Eq.~(\ref{waveorigin}) to obtain $a$ at the new time slice. Inserting
this value into the leading order in $r$ of Eq.~(\ref{origingamma})
and Eq.~(\ref{originU}), we calculate $\hat{\gamma}$ and $\hat{U}$ at the
two first grid points, which then allows us to start the marching algorithms
described above for the metric update.

In our numerical implementation we use a standard second order
integration for the double integrals in Eq.~(\ref{Deltanum}) and we
determine the conformal factor according to Eq.~(\ref{conformalfactor})
afterwards. Furthermore, to solve the hypersurface equations for the 
auxiliary variables $\tau$ and $\mu$ ($\mu$ is needed for the Bondi
mass, Eq.~(\ref{bondimass})), we also use a second order
discretization.  By simply reading off the metric values at
$\Scri^{+}$, it is finally possible to determine the Bondi mass and news.

\section{Code tests}
\label{tests}

In order to validate the accuracy of our numerical code we have performed
various tests aimed to check the different regimes of the implementation.

\subsection{Spherically symmetric tests}

In this section we describe tests of our code in dealing with
spherically symmetric stellar models.
In order to set up equilibrium models for relativistic stars we
solve the Tolman-Oppenheimer-Volkoff equations in outgoing null
coordinates~\cite{PaF99}
\begin{eqnarray}
\label{TOV1}
p_{,r} & = & \left( \frac{1}{2r} - \frac{1}{2Y}(1+8 \pi r^{2} p) \right) \rho h, \\
\label{TOV2}
Y_{,r} & = & 1 + 8 \pi r^{2} (p - \rho h),
\end{eqnarray}
where $Y=V e^{-2 \beta}$. Moreover, for the study of oscillations
of these spherical stellar models discussed below, we give some additional 
fluid perturbation. In order to obtain consistent initial data, we have to solve 
the hypersurface equations~(\ref{Bondiequation})-(\ref{Vhyp}) imposing the
normalization condition for the fluid. As we have presented results for a 
similar code elsewhere~\cite{SFP01}, we keep the presentation of this section
short.

\subsubsection{Stable relativistic stars}

First we study equilibrium models of relativistic
stars. In order to obtain initial data we solve the TOV
equations~(\ref{TOV1})-(\ref{TOV2}) for a polytropic EoS $p=K\rho^{\Gamma}$,
$K$ being the polytropic constant. With complete initial data at hand we evolve 
stable equilibrium configurations in time. Both, for finite grids which only 
cover the star, as well as for compactified grids, where we cover the star and 
its entire exterior spacetime, we are able to maintain the initial equilibrium 
profiles of the star during a time-dependent simulation, for times much longer
than the light-crossing time. Due to the discretization error, the
stars are exited to oscillate in their radial modes of pulsations
(see~\cite{SFP01} for more details). Deviations from equilibrium
converge to zero with increasing resolution with a second order convergence 
rate.

\subsubsection{Migration of an unstable relativistic star}

Following~\cite{FGIM01} we have checked the code on the dynamical
evolution of an unstable spherical star. In such a model the
sign of the truncation error of the numerical scheme controls the fate
of the evolution: the star may either expand or collapse. In our code
this sign is such that the unstable star ``migrates" to the stable
branch of the sequence of equilibrium models. In such a situation, the
rest-mass of the star has to be conserved throughout the
evolution. Despite being an academic problem this simulation
represents an important test of the accuracy and self-consistency of
the code in a highly dynamical situation.

\begin{figure}[t]
\centerline{\psfig{file=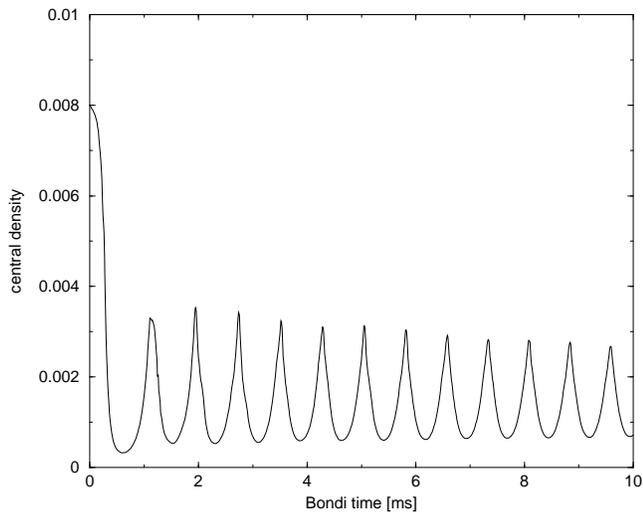,width=2.7in,angle=-90}}
\caption{Evolution of the central rest-mass
density during the migration of an unstable
relativistic star ($n=1, K=100, M=1.447 M_{\odot}, \rho_c=8.0 \times
10^{-3}; G=c=M_{\odot}=1$) to a stable model with the same
rest-mass. The central density of the (final) stable configuration
is $\rho_c=1.35\times 10^{-3}$. The evolution shows the expected
behavior. Since we are using a polytropic EoS, the amplitude of
the oscillations is essentially undamped for the evolution times
shown.}
\label{migration}
\end{figure}

As in~\cite{FGIM01} we have constructed a $n=1$ ($\Gamma=1+1/n=2$),
$K=100$ polytropic star with mass $M=1.447\;M_\odot$ and central
rest-mass density $\rho_c=8.0\times 10^{-3}$ (in units in
which $G=c=M_{\odot}=1$).
As the radius of the star strongly increases during the evolution, we
surround the star by a low density atmosphere. In order to
avoid numerical problems in these zones we reset the fluid
variables to their original atmosphere values once they have fallen
below a threshold value~(see~\cite{FMST98} for more details). 
This artificial resetting enables the star to expand and to contract. 
Despite the use of an atmosphere the energy conservation properties 
are well satisfied \cite{SFP01}. 

Fig.~\ref{migration} shows the evolution of the central density up to
a final time of $u_B=10$ ms. On a very short dynamical timescale the star
rapidly expands and its central rest-mass density drops well below its
initial value, less than $\rho_c=1.35\times 10^{-3}$, the central
rest-mass density of the stable model of the same rest-mass. During
the rapid decrease of the central density, the star acquires a large
radial momentum.  The star then enters a phase of large amplitude
radial oscillations around the stable equilibrium model. As
Fig.~\ref{migration} shows the code is able to accurately recover
(asymptotically) the expected values of the stable model. Furthermore,
the displayed evolution is completely similar to that obtained with an
independent fully three-dimensional code in Cartesian
coordinates~\cite{FGIM01}.

The evolution shown in Fig.~\ref{migration} allows to study large
amplitude oscillations of relativistic stars, which could
occur after a supernova core-collapse~\cite{dimmelmeier} or after an
accretion-induced collapse of a white dwarf.

\subsection{Tests beyond spherical symmetry}

To the best of our knowledge we do not know of any regular exact
(analytic) solution of the Einstein equations in axisymmetry with a 
non-vanishing perfect fluid matter field. However, there is an
exact solution for the vacuum equations which was already used to
check the vacuum code of~\cite{GPW94}. As we have generalized the
coordinate system, we use here the same solution to check our code.

In addition, to test the overall implementation of the code we examine
the global conservation properties which can be rigorously established 
due to the compactification of spacetime.

\subsubsection{Exact vacuum solution SIMPLE}

Following the work of~\cite{GPW94} the metric 
\begin{eqnarray}
e^{\gamma} & = &\frac{1}{2}(1+\Sigma), \\
e^{2 \beta} & = & \frac{(1+\Sigma)^{2}}{4 \Sigma}, \\
U & = & = - \frac{a^{2} g \cos \theta}{\Sigma}, \\ 
V & = & = \frac{r}{\Sigma}(2 a^2 g^2 - a^{2} r^{2} + 1),
\end{eqnarray}
is a solution of the vacuum field equations for
\begin{eqnarray}
\Sigma & = & \sqrt{1+a^{2} g^{2}}, \\
g & = & r \sin \theta,
\end{eqnarray}
where $a$ is a constant. Using this solution with a suitable L2-norm
to measure deviations we have checked that our metric solver is second 
order convergent.

\subsubsection{Global energy conservation test}

\begin{figure}[t]
\centerline{\psfig{file=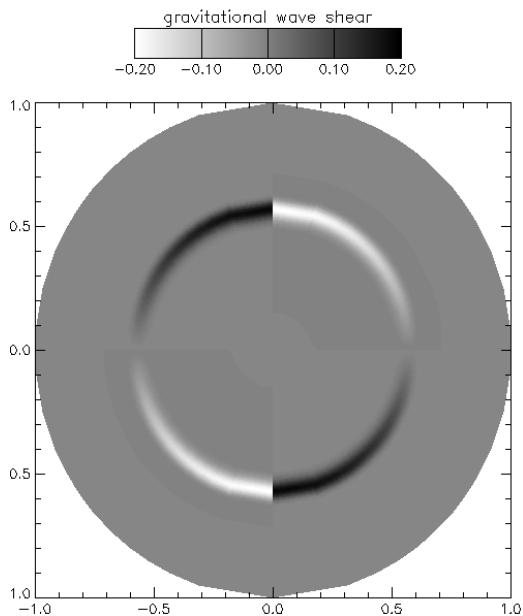,width=2.7in}}
\caption{Contour plot of the initial gravitational wave shear $\hat{\gamma}$,
Eq.~(\ref{pertg}). The axes labels denote the value of the radial
coordinate $x$ along the equator (horizontal axis) and the symmetry
axis (vertical axis), and they run from the origin of the coordinate system
at $x=0$ to future null infinity $\Scri^{+}$ at $x=1$. The
numerical domain comprises the half circle to the right of the
vertical line at $x=0$ (the symmetry axis), the left half is obtained
from axisymmetry. The gravitational wave shear perturbation is located
in a small ring at radius $r \approx 4$.\label{ctgplot}}
\end{figure}

In this section we focus on a global energy conservation
test. Starting with the equilibrium model of a neutron star (see next
section for more details), we use a strong gravitational wave to
perturb the star,
\begin{equation}
\label{pertg}
\hat{\gamma} = 0.2 \ e^{-3(r-4)^{2}} y.
\end{equation}

Such a large amplitude is not realistic, but we choose it to test our
numerical implementation in the nonlinear regime.
Fig.~\ref{ctgplot} shows a contour plot of the initial gravitational
wave shear. In what follows we use a radial grid with $r = 3 \frac{x}{1-x}$. 

\begin{figure}[t]
\centerline{\psfig{file=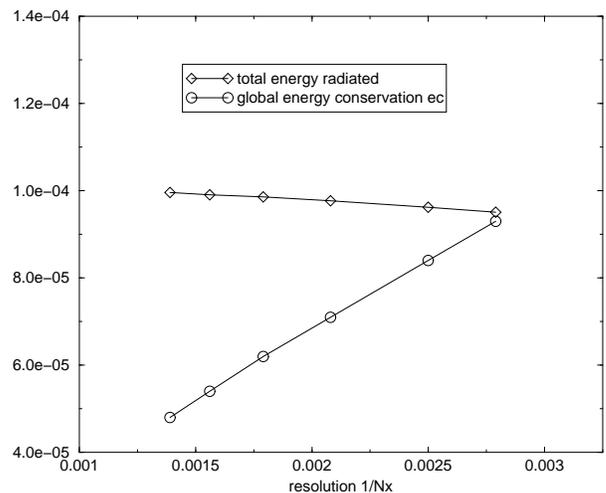,width=2.6in,angle=-90}}
\caption{Global energy conservation test for a neutron star and a
strong gravitational wave. Plotted are the deviation from global
energy conservation $ec$~(circles, see Eq.~(\ref{ec})) and
the total energy emitted by gravitational waves as a function of the grid 
resolution. The final integration time is $u_B = 2 \ 10^{-8}$~s, the total 
number of angular grid points $ 2 N_y  = 0.1 N_x$.}
\label{fig3}
\end{figure}

Fig.~\ref{fig3} shows the deviation from global energy conservation
(see Eq.~(\ref{ec})) as a function of the grid resolution (circles) and the 
total energy radiated away in gravitational waves. The deviations from exact 
energy conservation converge to zero which represents a
very severe global test for our numerical implementation.

\begin{figure}[t]
\centerline{\psfig{file=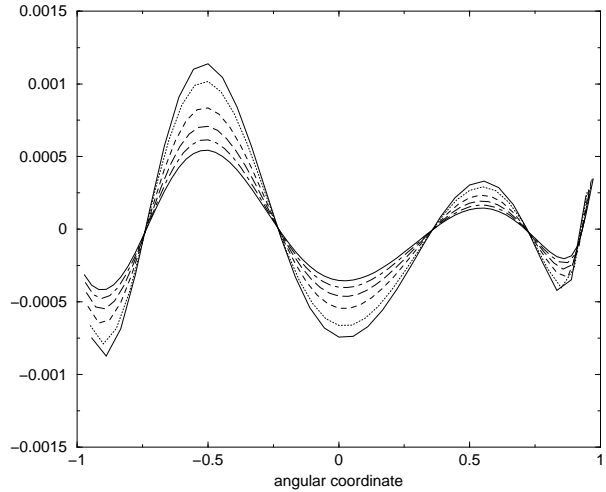,width=2.6in,angle=-90}}
\caption{Consistency check for the global norm Eq.~(\ref{normS}) at
$\Scri^+$. Plotted are deviations from zero as a function of the angular 
coordinate $y$ and for the same resolutions used in Fig.~\ref{fig3}. 
The errors decrease with resolution, the convergence rate is one.}
\label{fig4}
\end{figure}

There are two additional consistency conditions, which relate the
metric quantities on $\Scri^{+}$. These are~\cite{Pap93}
\begin{eqnarray}
\label{normS}
S - U_{,\theta} - U \ \cot\theta & = & 0,\\
\label{normg}
\gamma_{,u} + \frac{1}{2}e^{-2 \gamma} \sin\theta\left(e^{2 \gamma}
\frac{U}{\sin \theta}\right)_{,\theta} & = & 0.
\end{eqnarray}
Our code reproduces these conditions, the errors converging to zero with a
convergence rate of 1. Fig.~\ref{fig4} shows the deviation from zero for
the first condition, thus checking the leading term in the falloff behavior 
of the quantity $S$ at $\Scri^+$.

The obtained first-order convergence rate can be explained by the use
of a {\it total variation diminishing} HRSC scheme for the fluid evolution, 
which, although it is second-order accurate in smooth, monotonous parts of the flow,
reduces to first-order at local extrema, which are present in the
interior of the numerical domain in this test (see~\cite{FSK99} for alternative
{\it essentially nonoscillatory} schemes). Tests including
propagation and scattering off the origin of pure vacuum
gravitational fields yielded the expected second-order convergence.

\section{Pulsations of relativistic stars}
\label{pulsations}

Oscillations of neutron stars, although believed to emit only weak
gravitational waves (at least in the absence of instabilities) are
interesting sources for future gravitational wave detectors (for a
recent review see~\cite{KoA01}).

In this section we apply our code to the study of axisymmetric
pulsations of relativistic stars.
We focus our analysis on one specific stellar model. We
choose a relativistic polytrope $p = K \rho^{1 +\frac{1}{n}}$
with polytropic index $n=1$, polytropic constants $K=100$ and
central density $\rho_c = 1.28 \times 10^{-3}$ (in units in which
$c=M_{\odot}=G=1$). This model, which has a total mass of $M = 1.4
M_{\odot}$ has already been used in previous work~\cite{FSK99,FDGS01,FGIM01},
which allows us to compare our results for the radial frequencies and fixed 
background evolutions in axisymmetry.

\subsection{The perturbations}

In order to excite the radial oscillation modes of the star ($l=0$)
we perturb the equilibrium configuration using the
perturbation
\begin{eqnarray}
\delta \rho & = & A \rho_c \sin \left( \frac{\pi r^2}{R^2} \right), \\
\delta p & = & (1+\frac{1}{n}) p \frac{\delta \rho}{\rho},
\end{eqnarray}
where $A$ is the amplitude of the perturbation.

\begin{figure}[t]
\centerline{\psfig{file=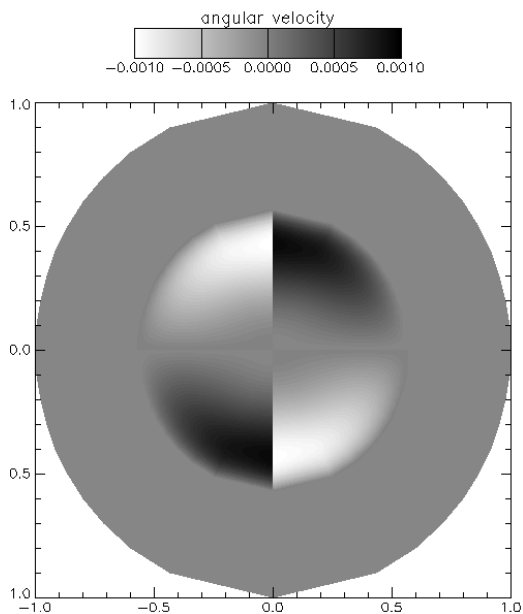,width=2.7in}}
\caption{Contour plot of the angular velocity perturbation $u_y$,
Eq.~(\ref{pertl2}). The axes labels denote the value of the radial
coordinate $x$ along the equator (horizontal axis) and the symmetry
axis (vertical axis), and run from the origin of the coordinate system
at $x=0$ to future null infinity $\Scri^{+}$ at $x=1$.
The star corresponds to the inner circle, where the
velocity perturbations are non-zero, with the north pole
above.\label{ctplot}}
\end{figure}

Additionally, to excite the $l=1,2$ non-radial modes we perturb the angular velocity 
component according to 
\begin{eqnarray}
u_y & = & A \sin \left( \frac{\pi r^2}{R^2} \right), \\
\label{pertl2}
u_y & = & A \sin \left( \frac{\pi r^2}{R^2} \right) \cos \theta,
\end{eqnarray}
respectively. Fig.~\ref{ctplot} shows the initial setup and a contour plot of the
perturbation given by Eq.~(\ref{pertl2}).
 
Following~\cite{FSK99,FDGS01}, in order to determine the different 
oscillation modes, we analyze the time evolution of different (fluid 
and metric) variables at a fixed coordinate location. We have checked that the 
frequencies of the oscillation modes are largely independent of the 
specific location. Hence, for the results presented here, we have restricted 
ourselves to extracting those frequencies at coordinates 
$(x,y)=(\frac{\tilde{N}_x}{2} \Delta x, \frac{N_y}{2} \Delta
y)$, where $\tilde{N}_x$ denotes the number of radial zones covering the
star. In addition we use a radial coordinate
\begin{equation}
r = 15 \frac{x}{1-x^{4}},
\end{equation}
and we choose a perturbation amplitude of $A=10^{-3}$. We focus on the
time evolution of the radial velocity $u^x$ for the extraction of
radial modes and on the angular velocity $u_y$ for the extraction of
non-radial modes. The calculation of mode frequencies follows the work
of~\cite{FDGS01}, i.e. we determine the zeros of the first
derivative of the Fourier transform. These zeros correspond to 
maxima in the Fourier transform which are associated with the
excited modes of oscillation.

\subsection{Fixed background evolutions}

\begin{table}[t]
\caption{
Mode frequencies obtained in the Cowling approximation for a
relativistic polytrope with $K=100$, $n=1$ and central density $\rho_c
= 1.28 \times 10^{-3}$ in units in which $c=G=M_{\odot}=1$. The
first column labels the different modes. The second column
shows the frequencies obtained with our code, the third column
the results obtained with a different nonlinear code based on Cauchy
slices~\cite{FDGS01}. The fourth column indicates the frequencies
obtained from a linear perturbation code~\cite{FSK99} for the
quadrupolar modes. The last column shows the relative difference
between the present code and the results of~\cite{FDGS01} in percent.}
\medskip
\begin{ruledtabular}
\begin{tabular}{ccccc}
Mode & Present Code & Code~\cite{FDGS01} & Perturbation~\cite{FSK99} & Difference  \\
     &  (kHz)       &   (kHz)            &   (kHz)                   &  (per cent) \\
\hline \hline
$F$ & 2.690 & 2.706 & - & 0.59
\\
$H_{1}$ & 4.636 & 4.547 & - & 1.96
\\
$H_{2}$ & 6.532 & 6.320 & - & 3.35
\\
$H_{3}$ & 8.418 & 8.153 & - & 3.25
\\
${^{1}f}$ & 1.388 & 1.335 & - & 3.97
\\
${^{1}p_{1}}$ & 3.504 & 3.473 & - & 0.89
\\
${^{1}p_{2}}$ & 5.510 & 5.335 & - & 3.28
\\
${^{1}p_{3}}$ & 7.400 & 7.136 & - & 3.70
\\
${^{2}f}$ & 1.871 & 1.852 & 1.884 & 1.03
\\
${^{2}p_{1}}$ & 4.143 & 4.100 & 4.110 & 1.05
\\
${^{2}p_{2}}$ & 6.135 & 6.019 & 6.035 & 1.93
\\
${^{2}p_{3}}$ & 8.087 & 7.867 & 7.873 & 2.80
\\
\end{tabular}
\end{ruledtabular}
\label{Cowling}
\end{table}

We start presenting the results for the mode frequencies of the
above stellar model obtained in evolutions in which we fix the background 
geometry of the spacetime (i.e. we adopt the so-called Cowling approximation).  
We compare the results to the literature, thus testing the 
hydrodynamics solver of our code.

Table~\ref{Cowling} shows the results for the fundamental radial mode ($F$)
and the first three overtones ($H_1-H_3$) and the $f$ and $p$ modes for both, 
the $l=1$ and $l=2$ perturbation. Our results are in good agreement with both,
a different nonlinear code~\cite{FDGS01} and an independent linear code based
upon perturbation theory~\cite{FSK99}. There are several reasons which explain 
the observed differences: First, the grid resolution used in our simulations is 
rather low ($451 \times 21$ grid points in $x$ and $y$, respectively) compared
to the finer grids used in~\cite{FDGS01} ($200 \times 80$), especially
in the angular direction. For the current setup - in contrast
to~\cite{FDGS01} - only about one half of the radial 
zones is used to cover the star. We note that we use this choice here for 
comparisons with the results of the next section, where we also have to
resolve gravitational waves in the exterior spacetime. Second, we have not 
implemented an atmosphere surrounding the star in a few zones as in~\cite{FDGS01},
which enables the star to radially contract and expand. 
Therefore, the surface of the star may be too rigid, which may affect the mode
frequencies slightly. The numerical implementation of such an atmosphere, however, 
requires a problem-dependent adaptation of its parameters.
Third, and perhaps most importantly, as described in 
Sec.~\ref{sec:fluid} we use a second order reconstruction scheme at the cell 
interfaces, and not the third order PPM scheme used in~\cite{FDGS01}. It is worth 
emphasizing the importance of using high-order schemes for the hydrodynamics in order 
to improve the frequency identification (see related discussion in~\cite{FSK99}).
Nevertheless, for our only purpose of assessing the validity of the code we think 
that the overall agreement found is satisfactory.

\begin{figure}[t]
\hspace{-0.38cm}
\centerline{\psfig{file=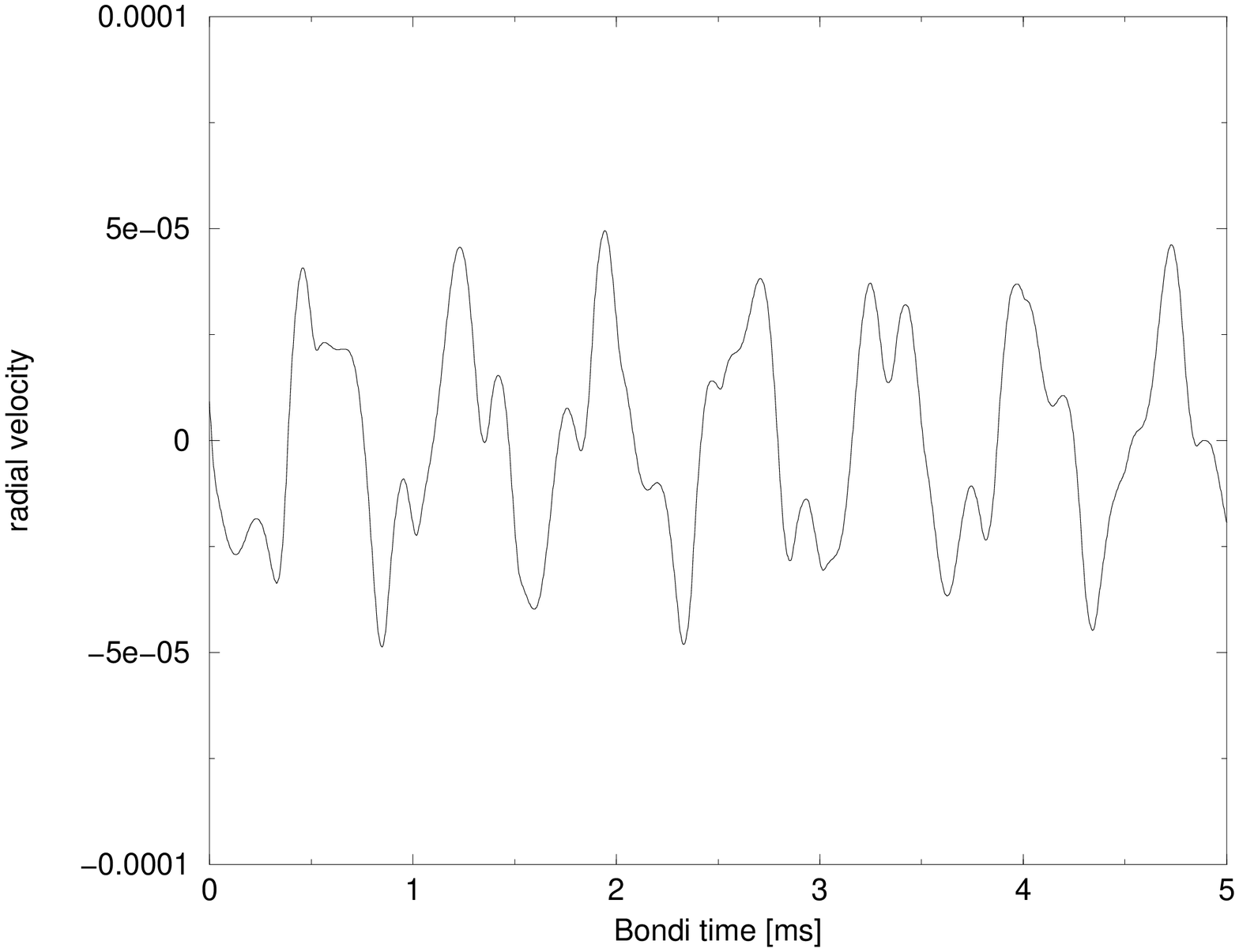,width=2.6in,angle=-90}}
\hspace{+0.38cm}
\centerline{\psfig{file=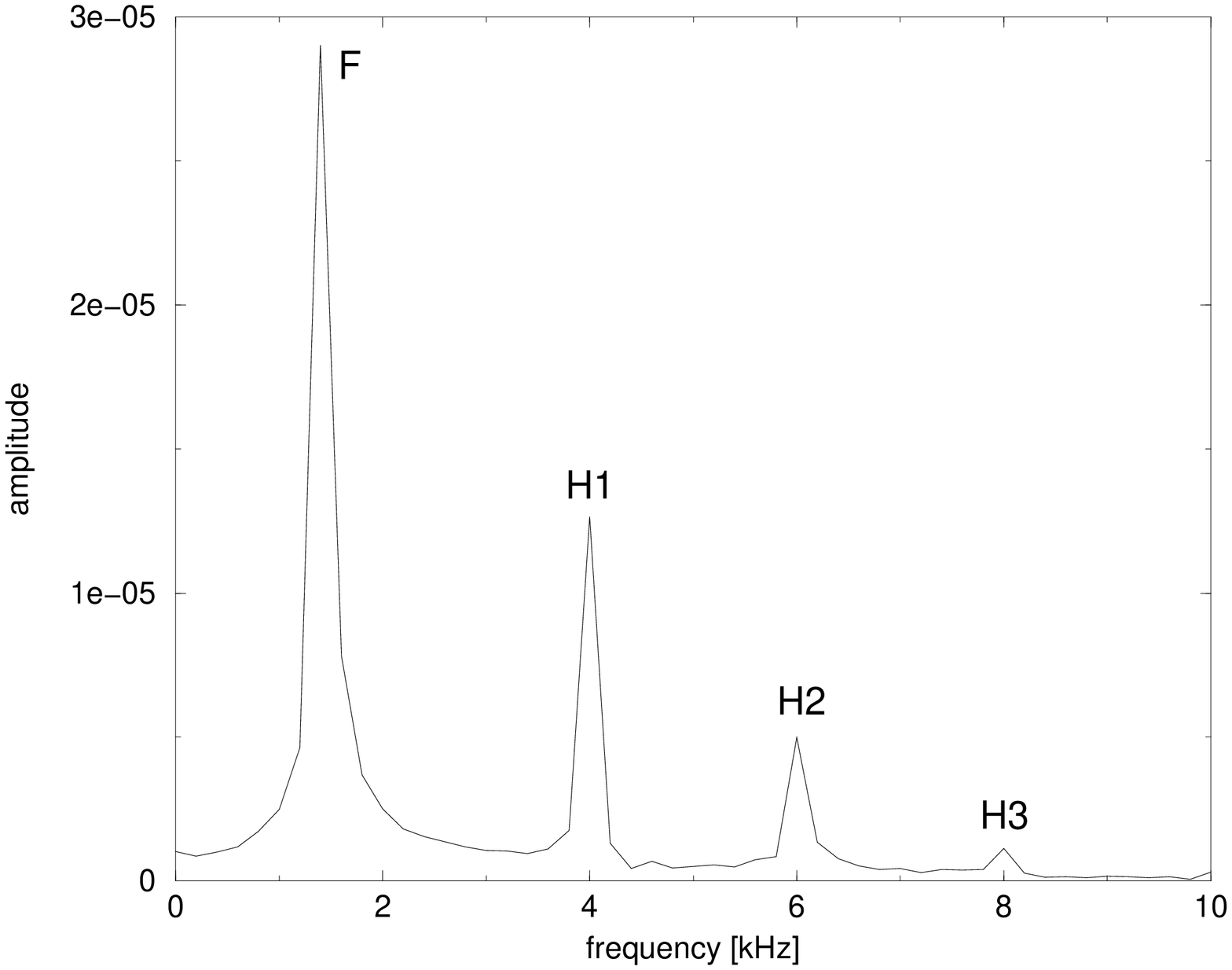,width=2.6in,angle=-90}}
\caption{
Pulsations of a $K=100, n=1, \rho_c = 1.28 \times 10^{-3}$ 
polytrope ($c=G=M_{\odot}=1$). 
Top panel: Time evolution of the radial velocity $u^{x}$ for the 
$l=0$ perturbation. Bottom panel: Fourier transform of the radial 
velocity. We have labeled the different oscillation modes, $F$ 
(fundamental) and $H_1-H_3$ (first three overtones).
\label{l0}
}
\end{figure}

\subsection{Metric-fluid coupled evolutions}

In this section we extract the mode frequencies of the above stellar model
from fully coupled evolutions of the fluid and the geometry.

\begin{figure}[t]
\hspace{-0.38cm}
\centerline{\psfig{file=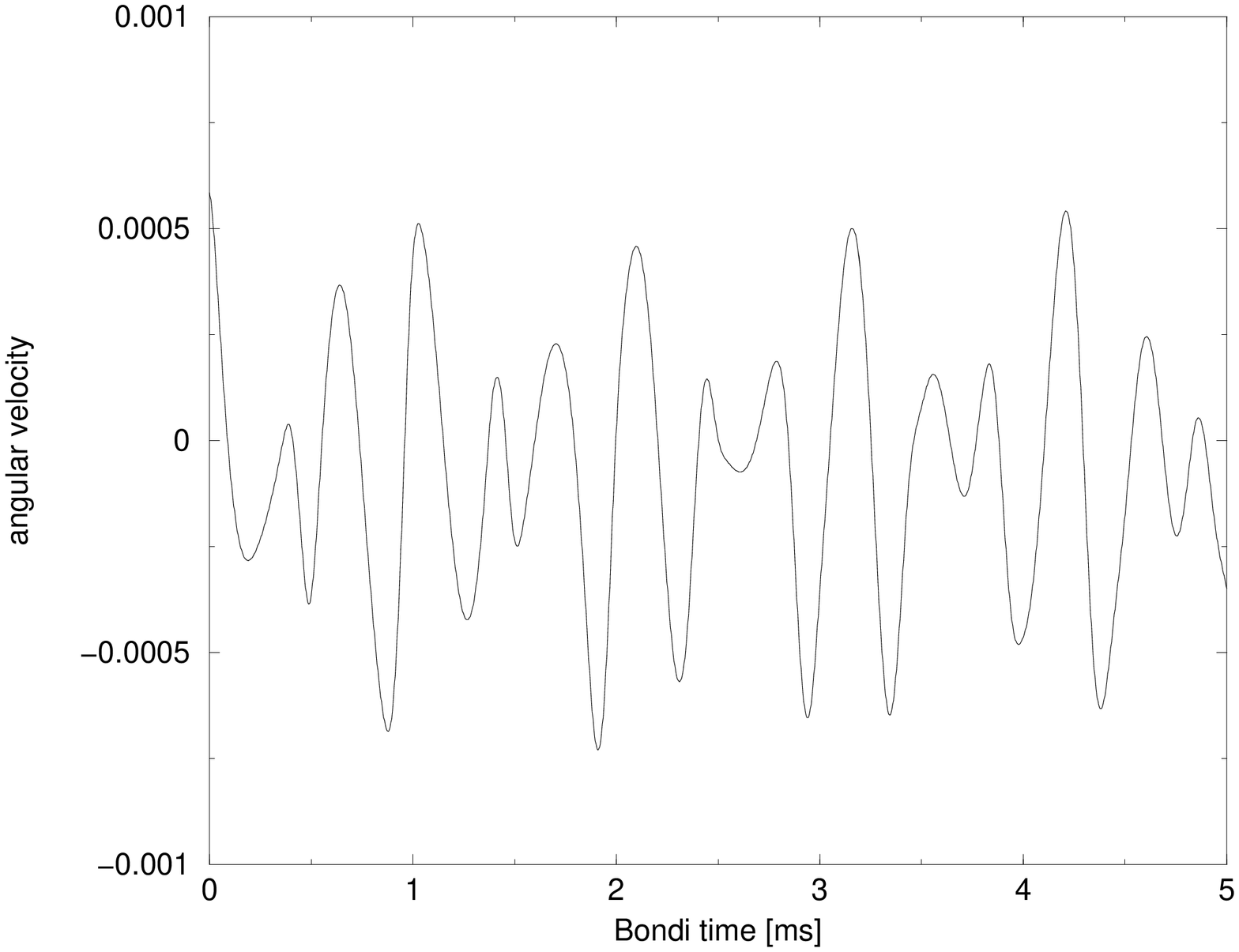,width=2.6in,angle=-90}}
\hspace{+0.38cm}
\centerline{\psfig{file=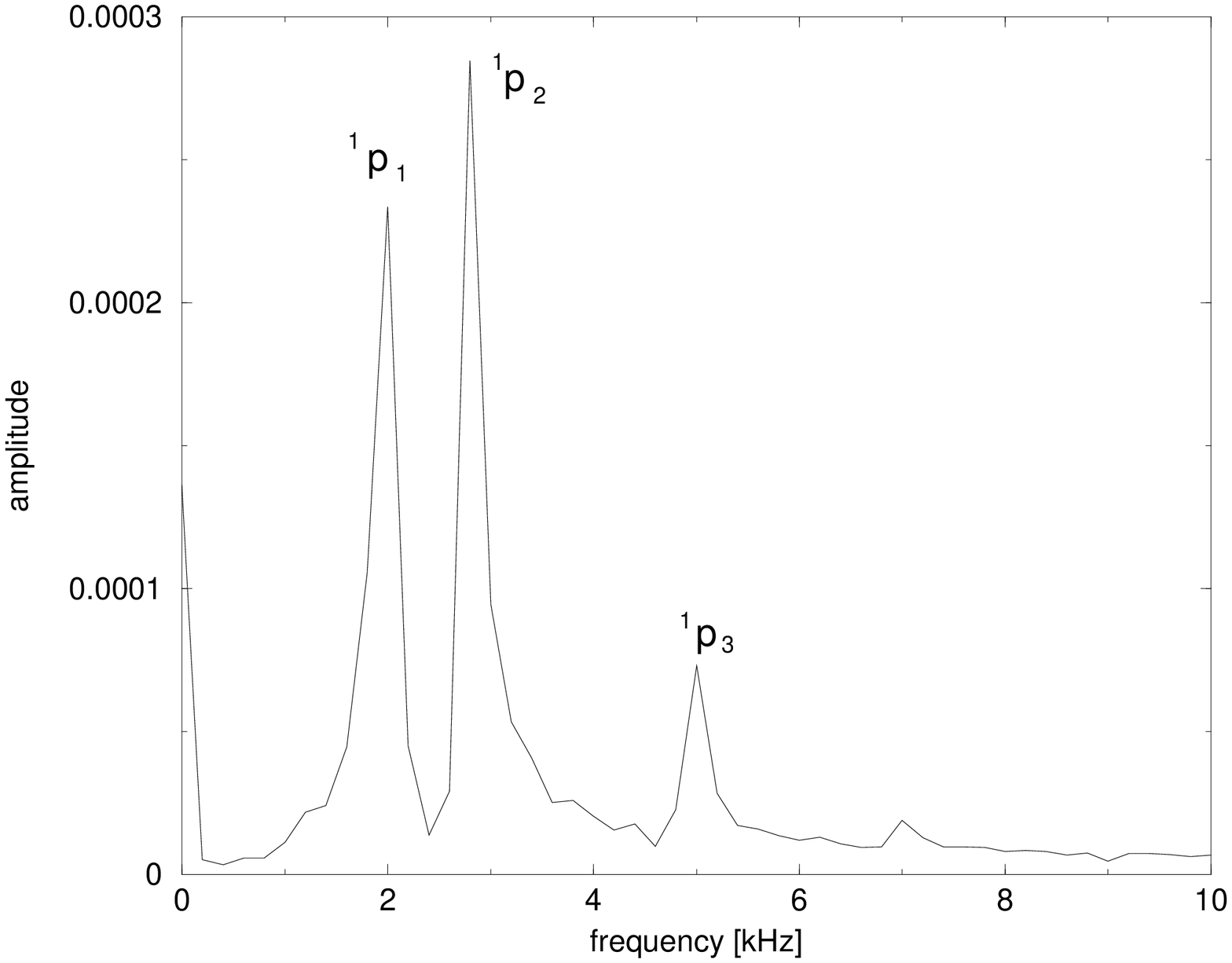,width=2.6in,angle=-90}}
\caption{
Pulsations of a $K=100, n=1, \rho_c = 1.28 \times 10^{-3}$ 
polytrope ($c=G=M_{\odot}=1$). 
Top panel: Time evolution of the angular velocity
$u_{y}$ for the $l=1$ perturbation.
Bottom panel: Fourier transform of the angular velocity.\label{l1}}
\end{figure}

\begin{figure}[t]
\hspace{-0.38cm}
\centerline{\psfig{file=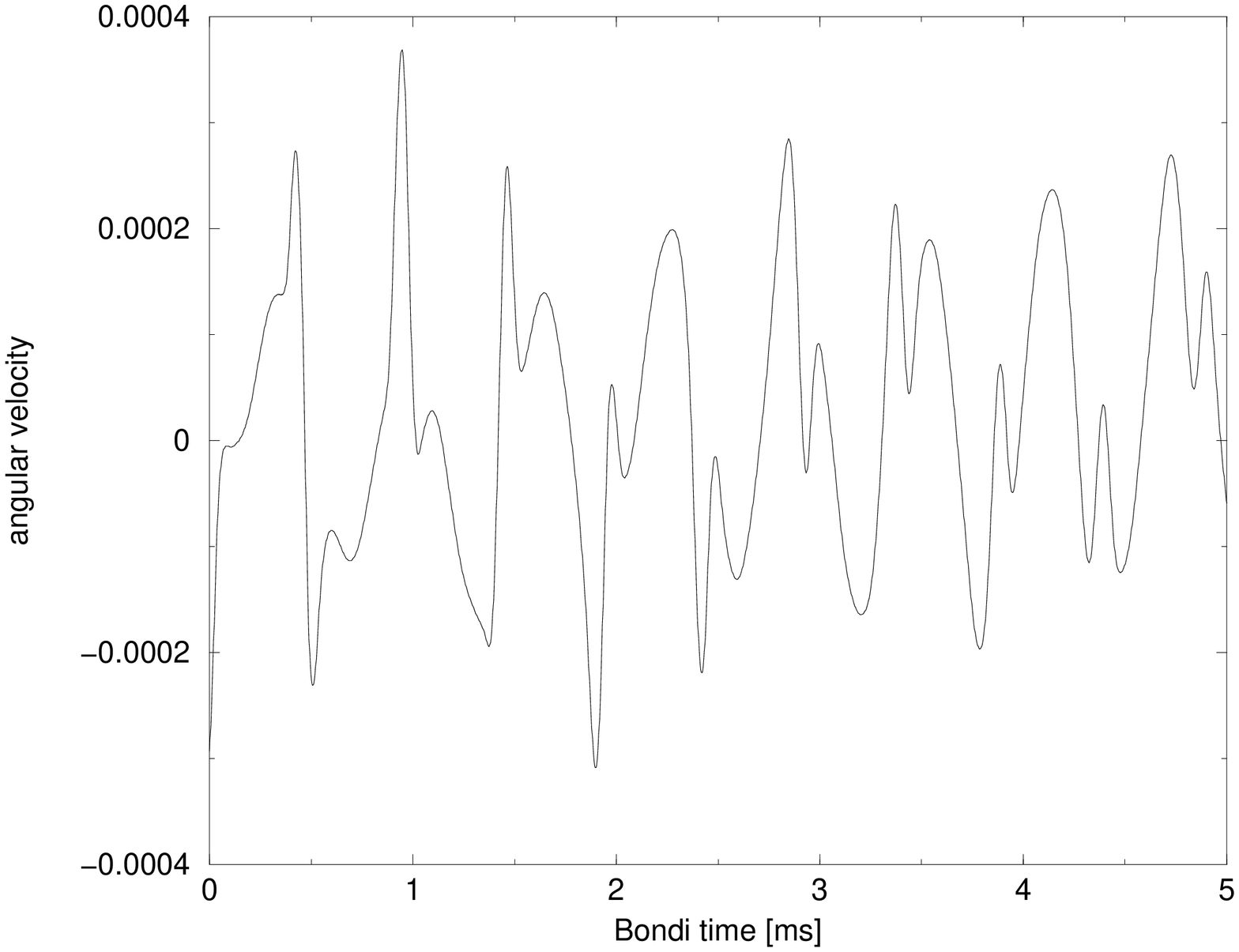,width=2.6in,angle=-90}}
\hspace{+0.38cm}
\centerline{\psfig{file=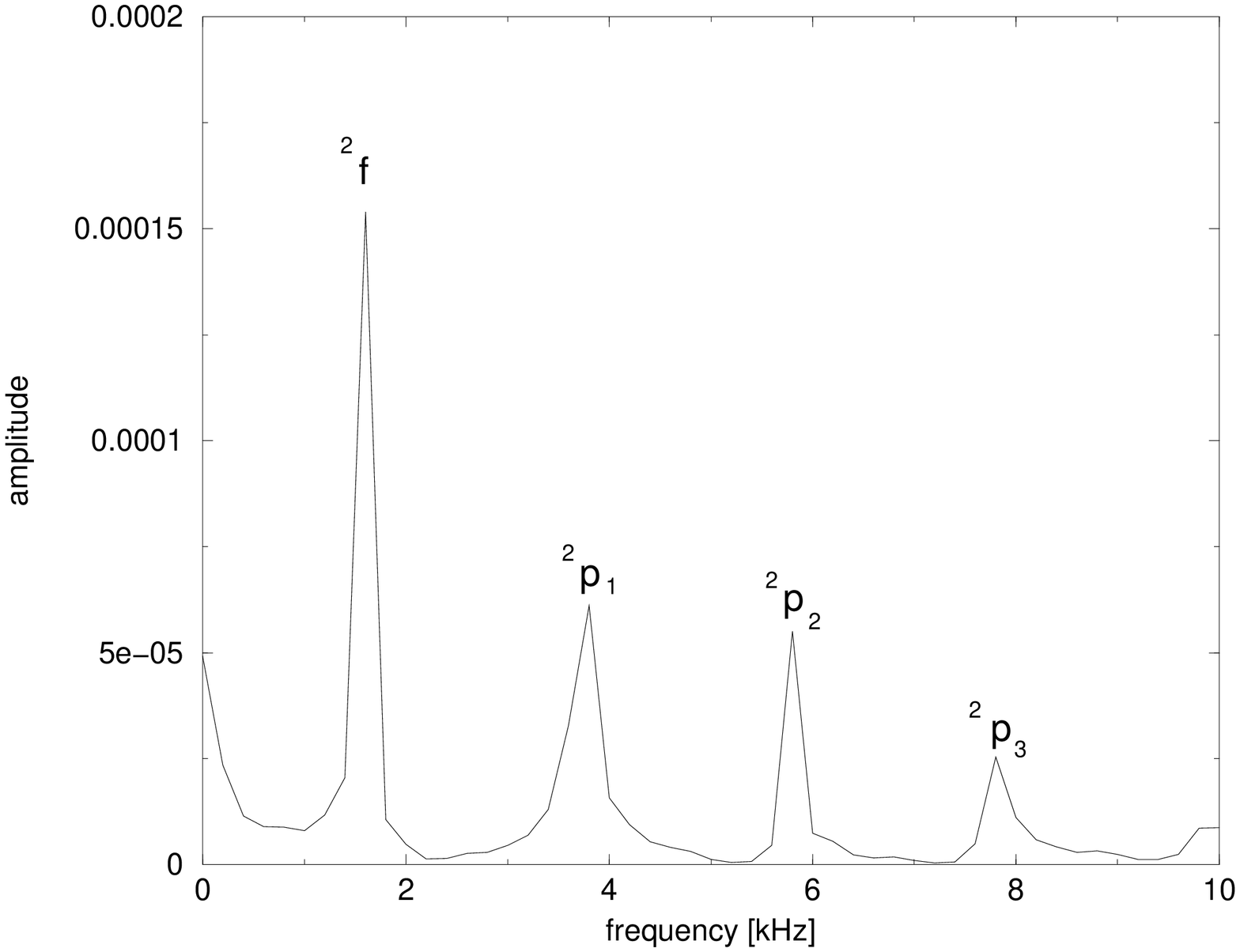,width=2.6in,angle=-90}}
\caption{
Pulsations of a $K=100, n=1, \rho_c = 1.28 \times 10^{-3}$ 
polytrope ($c=G=M_{\odot}=1$). 
Top panel: Time evolution of the angular velocity
$u_{y}$ for the $l=2$ perturbation.
Bottom panel: Fourier transform of the angular velocity.\label{l2}}
\end{figure}

Fig.~\ref{l0} shows the (Bondi) time evolution of the radial velocity
$u^{x}$ for the $l=0$ perturbation (top panel), as well as the Fourier transform
of this quantity. In the same way, Figs.~\ref{l1} and~\ref{l2}
display the angular velocity evolution $u_y$ and the Fourier transform
for the $l=1$ and $l=2$ perturbation, respectively.

The final evolution time in Figs.~\ref{l0}-\ref{l2} corresponds to
5 ms. The distinctive oscillatory pattern depicted in these figures 
is mainly a superposition of the lowest-order normal modes of the fluid. 
The high-frequency modes are usually damped fast by the intrinsic
viscosity of the numerical schemes and at late times the star mostly pulsates
in its lowest frequency modes.

We summarize our results on the mode frequencies in
Table~\ref{coupled}. Note that due to the conservation of linear momentum
the ${^{1}f}$ mode does not exist (see Fig.~\ref{l1})~\cite{FDGS01}.
As in the Cowling simulations of the
previous section we also find now good agreement when
comparing to results of an independent nonlinear code~\cite{FGIM01}
and to results of linear perturbation theory~\cite{FGIM01,Ko01}.
The reasons mentioned in the above section for the observed
discrepancies are still valid here, together with the new
source of error introduced by the metric evolution.

\begin{table}[t]
\caption{ Mode frequencies obtained in the coupled evolution for the
relativistic polytrope with $K=100$, $n=1$ and central density $\rho_c
= 1.28 \times 10^{-3}$ in units in which $c=G=M_{\odot}=1$. 
The first column labels the different modes.
The second column
shows the frequencies obtained with our code and the third column
shows the results obtained from linear perturbation
theory~\cite{FGIM01,Ko01}. The last column shows the
deviations in percent.}
\medskip
\begin{ruledtabular}
\begin{tabular}{cccc}
Mode & Present Code & Perturbation~\cite{FGIM01}, \cite{Ko01}& Difference \\
     &  (kHz)       &   (kHz)                                &  (per cent) \\
\hline \hline
$F$ & 1.422 & 1.442 & 1.38
\\
$H_{1}$ & 3.993 & 3.955 & 0.96
\\
$H_{2}$ & 6.021 & 5.916 & 1.77
\\
$H_{3}$ & 7.968 & 7.776 & 2.46
\\
${^{1}p_{1}}$ & 1.951 & - & -
\\
${^{1}p_{2}}$ & 2.844 & - & -
\\
${^{1}p_{3}}$ & 5.019 & - & -
\\
${^{2}f}$ & 1.587 & 1.579 & 0.51
\\
${^{2}p_{1}}$ & 3.748 & 3.710 & 1.02
\\
${^{2}p_{2}}$ & 5.811 & 5.689 & 2.14
\\
${^{2}p_{3}}$ & 7.848 & 7.580 & 3.54
\\
\end{tabular}
\end{ruledtabular}
\label{coupled}
\end{table}

\subsection{Gravitational waveform}

\begin{figure}[t]
\centerline{\psfig{file=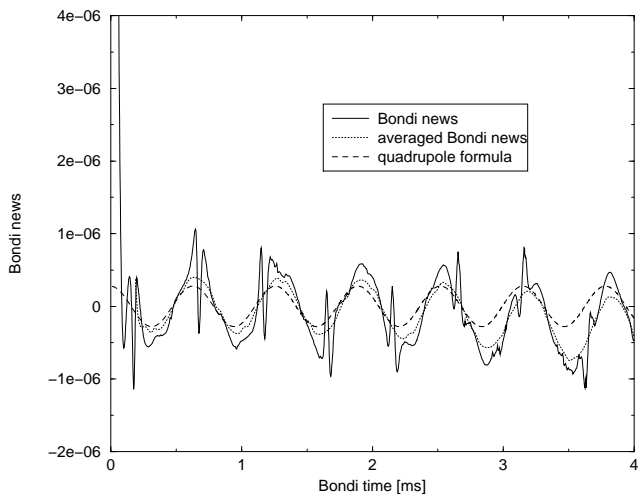,width=2.6in,angle=-90}}
\caption{News function as a function of observer time at
infinity at the equator. The (unphysical) initial signal is due to the
gravitational wave content in the initial data. The main oscillation 
frequency corresponds to the ${^{2}f}$ mode with other
frequencies overlayed. The dotted curve shows a suitable average 
of the numerical Bondi news smearing out higher frequencies. 
The dashed line corresponds to the estimated amplitude of the
${^{2}f}$ mode oscillation which is extracted using the
quadrupole formula. See text for a detailed discussion.\label{news}}
\end{figure}

To end the validation of our code we study the gravitational wave signal from the
simulations of the above section concerning the $l=2$ perturbation. In Fig.~\ref{news} 
we plot the Bondi news function at
the equator $N(y_B=0)$ as a function of the observer time. Due to the
equatorial plane symmetry inherent to the perturbation, which is
conserved during the evolution, we have $y_B(y=0)=0$, which enables us
to directly plot $N(y=0)$, thus avoiding suitable interpolations for
the wave extraction.

In order to estimate the amplitude of the signal in Fig.~\ref{news} we
have also computed the gravitational wave emission due to the quadrupole
formula. In the light-cone approach, the {\it quadrupole news} takes 
the form \cite{Win87}
\begin{equation}
\label{qnews}
N_{0} = \dddot{Q}.
\end{equation}
The relevant quadrupole moment in axisymmetry reads
\begin{equation}
\label{quad}
Q = \pi \sin^{2} \theta \int_{0}^{R} dr' \int_{0}^{\pi} \sin \theta'  d\theta' r'^{4}
\rho (1+ \epsilon)
(\frac{3}{2} \cos^{2} \theta' - \frac{1}{2}).
\end{equation}
We have included the factor $(1+\epsilon)$ in the formula to account
for the relativistic internal energy.
In the numerical calculation of $Q$ we equate the radial and angular
coordinates $r'$ and $\theta'$ with the radial and angular coordinates
$r$ and $\theta$. It is well known~\cite{Fin89} that the three time
derivatives in Eq.~(\ref{qnews}) can cause severe numerical error. In
order to avoid this problem we fit a cosine to the time evolution
of the quadrupole moment $Q$
obtained in the numerical simulation. The dashed line in
Fig.~\ref{news} shows the third time derivative of this curve, 
where the time derivatives are taken with repect to Bondi time.

By evaluating the contributions of the gravitational wave signal in
Fig.~\ref{news}, we see that the dominant contribution originates from the
${^{2}f}$ mode. The frequency we extract from the waveform is 1.57
kHz, in good agreement with the result shown in Table~\ref{coupled}.
There are additional contributions to the gravitational wave signal. 
The excitation of the ${^{1} p_{1}}$ mode is created at the outer boundary 
of the fluid, where we do not allow the star to radially contract 
or to expand.

Comparing the amplitude of the Bondi news and the quadrupole
news for the ${^{2}f}$ mode we stress that the amplitude roughly
agree. We cannot exclude that the differences are mainly due to numerical
errors. Due to the different contributions of the time derivative 
for $\gamma$ and gauge terms the calculation of $N$ is a difficult
task ~\cite{IWW84}. In addition, we note that the total energy
radiated away for our setup corresponds only to a tiny fraction of the
total mass of the spacetime. More precisely, whereas the total spacetime 
mass is $1.4 M_\odot$, the total mass radiated is only $2.8 \times 10^{-9}
M_{\odot}$, which is smaller than the typical numerical errors in the
determination of the Bondi mass. 
Moreover, most of this radiation is a consequence of 
the initial gravitational wave content and is radiated away in between 
$u_B=0$ ms and $u_B=0.1$ ms.  Additional differences in the curves for 
the Bondi news and the quadrupole news in Fig.~\ref{news} might be created 
by the rough estimate of the quadrupole Q in Eq.~(\ref{quad}) which does 
not take into account the curvature of spacetime.   

At late times (after about 3 ms), the Bondi news does not strictly
oscillate around zero. This numerical effect can be weakened 
using more sophisticated numerical methods for the
fluid update (e.g. third-order cell reconstruction procedures). 
We note that similar drifts are reported in~\cite{FSK99,FGIM01}
in time evolutions of equilibrium models using a perfect fluid
EoS.  For the results shown in Fig.~\ref{news}, we have
used a polytropic EoS during the fluid evolution. This is legitimate 
as the effect of heating is negligible for our stellar object close 
to equilibrium. Finally, we note that the maximum deviation from global 
energy conservation, Eq.~(\ref{ec}), in this simulation is $2.5
\times 10^{-5} M_{\odot}$, which is very adequate for the long
integration time.

\section{Conclusion}

We have presented numerical algorithms to solve the coupled
Einstein-perfect fluid system in axisymmetry. Our approach is based 
upon the characteristic formulation of general relativity in which
spacetime is foliated with a family of outgoing light cones
emanating from a regular center. Due to a suitable compactification 
of the spacetime future null infinity is part of our finite numerical 
grid where we unambiguously extract gravitational waves.    

Applying the nonlinear, fully relativistic code to studies 
of neutron stars modeled as polytropes, it has passed several tests, 
aimed at testing both the fluid evolution as well as the metric solver in the 
nonlinear regime. The code can accurately maintain long-term stability of
relativistic stars and we have applied it to the study of stellar
pulsations. We have extracted the frequencies of different non-radial 
fluid modes and the gravitational wave signal arising in time evolutions 
of perturbed stellar configurations. Applications of the present code in 
the computation of the gravitational waveforms emitted in axisymmetric core 
collapse situations will be presented elsewhere.

\section{Acknowledgements}

It is a pleasure to thank Kostas Kokkotas, who calculated mode
frequencies for us using his linear perturbation code. 
We also would like to thank Nick Stergioulas and Luis Lehner 
for helpful comments and discussions. This work has been supported in part
by the EU Programme 'Improving the Human Research Potential and the
Socio-Economic Knowledge Base', (Research Training Network Contract
HPRN-CT-2000-00137). P.P. acknowledges support from the Nuffield
Foundation (award NAL/00405/G). J.A.F acknowledges support from a
Marie Curie fellowship from the European Union (MCFI-2001-00032).

\end{document}